\documentclass[reprint,twocolumn,amsmath,amssymb,aps]{revtex4-2}
\usepackage{graphicx}
\usepackage{dcolumn}
\usepackage{bm}
\usepackage{physics}
\usepackage[normalem]{ulem}
\usepackage[dvipsnames]{xcolor}
\usepackage{nicefrac}
\usepackage{lineno}

\DeclareMathOperator{\arctanh}{arctanh}

\definecolor{myblue}{RGB}{40,100,200}
\usepackage[
    colorlinks=true,
    linkcolor=myblue,
    citecolor=myblue,
    urlcolor=myblue,
    pdfborder={0 0 0}
]{hyperref}

\begin{document}

\title{On Optimal Measurement-State Preparation via Geometric Transport of the Squeezing Ellipse}

\author{Vsevolod Salakhutdinov$^{1}$}
\email[corresponding author: ]{vsevolod.salakhutdinov@mpl.mpg.de}
\author{Nikolay Kalinin$^{1}$}

\affiliation{$^1$Max Planck Institute for the Science of Light, 
Erlangen, Germany}

\date{\today}

\begin{abstract}
Preparation of optimal measurement states is a key requirement in quantum metrology utilizing squeezed states. Besides positioning the mean state, such preparation requires proper alignment of the squeezing ellipse. We discuss a geometric framework that transforms an initially misaligned squeezed input into a measurement-optimal state in SU(2)-symmetric systems. Within this framework, the orientation of the squeezing ellipse constitutes an additional geometric degree of freedom and evolves as the mean state follows a controlled trajectory on the unit sphere. The resulting rotation of the ellipse depends on the chosen trajectory and is linked to the geometric phase through the evolution of the local basis. The discussed framework is applicable to different physical platforms admitting an effective SU(2) description. As a particular example, we consider polarization-squeezed light and outline a possible implementation using a continuously varying birefringent element.
\end{abstract}

\maketitle

\section*{\textbf{Introduction}}
Control of quantum systems is central to modern experimental quantum science, particularly in the preparation of states that enhance measurement precision. In quantum metrology, this objective is directly linked to surpassing classical resolution limits through the exploitation of nonclassical resources \cite{Degen2017,Pezze2018}.

Squeezed states constitute one of the primary mechanisms for achieving such enhancement \cite{Walls1983, Xu2019}. Their anisotropic uncertainty distribution, represented by an elliptic-contoured Wigner function in phase space, enables noise suppression along one quadrature at the expense of increased fluctuations in the conjugate direction, consistent with the Heisenberg uncertainty principle. Such states arise in a variety of platforms, including collective spins in two-level atomic ensembles \cite{Wineland1992, Kitagawa1993, Ma2011, Sinatra2022}, polarization \cite{Grangier1987,Chirkin1993,Bowen2002,Korolkova2002,Heersink2005} and spatially polarized~\cite{Lassen2009,Guo2017} states of light and magnons \cite{Kamra2016,Chauhan2026,Weng2026}. In all such settings, the metrological advantage critically depends on aligning the squeezed quadrature with the measurement axis. This condition is encoded in the orientation of the squeezing uncertainty ellipse, rendering the measurement-optimization problem intrinsically geometric. 

Squeezed states significantly enhance interferometric sensitivity and play a central role in precision phase estimation. In interferometric settings, the evolution of quantum states is naturally associated with phase accumulation. Besides the familiar dynamical phase (see, e.g., \cite{Klyshko1993} and Sec.~6.6 in \cite{Weinberg2015}), quantum systems may acquire an additional geometric contribution. It is determined solely by the trajectory traced in parameter space and is known as the geometric phase~\cite{Berry1984,Simon1983,Aharonov1987}. It appears as a global phase factor acquired upon cyclic evolution, $|\phi_{\rm end}\rangle = e^{i \xi}|\phi_{\rm start}\rangle$,
and has been exploited in interferometric and precision-sensing applications~\cite{Maclaurin2012,Jisha2021,Zhou2025}. However, the conventional formulation describes the evolution of quantum states viewed as rays in Hilbert space and therefore captures only the global phase of the state vector~\cite{Klyshko1993,Cisowski2022}. The presence of squeezing introduces an additional anisotropic uncertainty structure that is not fully characterized by this phase alone: the orientation of the squeezing ellipse constitutes a distinct geometric degree of freedom that must be transported and controlled together with the state vector. The geometric significance of fluctuation orientation can be illustrated by the net SO(3) rotation of the fluctuation ellipsoid resulting from parallel transport along closed loops of the spin vector within the Bloch ball, highlighting a deep connection between the geometry of real space and that of quantum-state space~\cite{Bharath2018}. Thus, turning squeezing-orientation transport into a constructive tool for metrological state preparation presents a related but distinct challenge that can be addressed through a geometric formulation.

In phase-sensitive quantum metrology, optimizing measurement sensitivity involves jointly accounting for the state evolution and the orientation of its squeezed uncertainty. While recent work has explored how a given quantum state should evolve during measurement to maximize its sensitivity, see, e.g.,~\cite{Reilly2023,Lecamwasam2024}, the present work instead addresses the complementary task of preparing and transforming a squeezed input state for optimal measurement. This preparation requires proper alignment of the squeezing ellipse with the measurement configuration to maximize sensitivity to a small phase shift $\delta\theta$. Motivated by previous experimental demonstrations of such alignment based on discrete polarization transformations \cite{Kalinin2023,Kalinin2023a}, we focus on developing a unified geometric framework for preparing measurement-optimal states, formulated in terms of continuous SO(3) transformations and exemplified by polarization-squeezed light on the Poincar\'{e} sphere. Within this framework, the preparation problem is formulated as the joint control of the mean-state trajectory and the squeezing orientation. The squeezing ellipse is transported by the local SO(3) transformations governing the mean-state evolution, while its orientation is evaluated relative to the evolving local tangent-plane basis. The local-basis contribution is related to the integral defining the enclosed solid angle, which also underlies the geometric-phase description. This description establishes a geometric relation between the chosen transformation and the resulting squeezing orientation, enabling the construction of trajectories that satisfy prescribed final alignment conditions, while the same trajectory geometry also governs the associated geometric-phase accumulation. By connecting geometric trajectory design with experimentally accessible control parameters, the resulting framework provides a physically transparent approach to achieving prescribed squeezing-ellipse alignment.

The paper first introduces the geometric framework and then discusses its implementation for polarization-squeezed light and possible extensions to collective-spin systems and spatial optical modes.

\section*{Geometric Framework}
\begin{figure}[t!]
    \centering
    \includegraphics[width=0.95\linewidth]{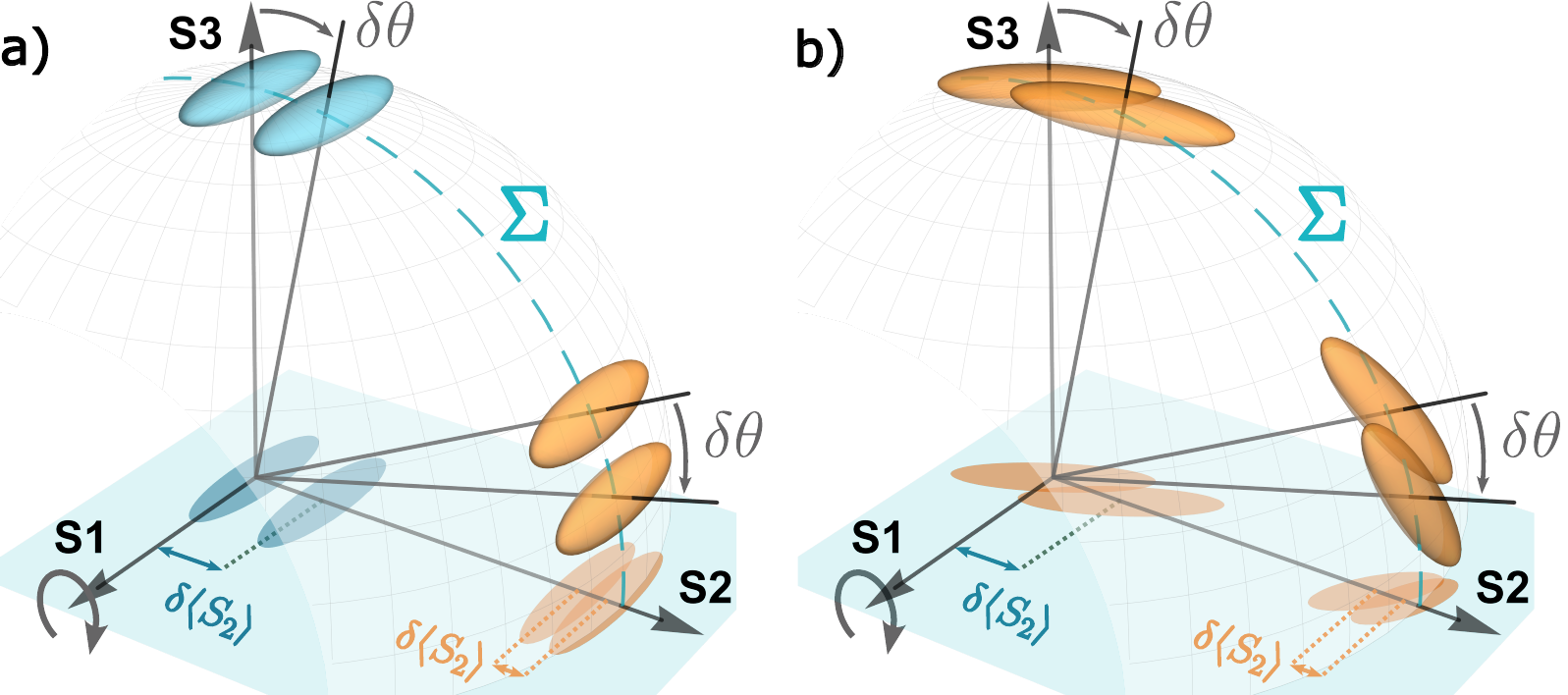}
    \caption{\textbf{Illustration of the optimal measurement state definition.} Initial states under different starting conditions evolve into final states displaced by a small phase variation $\delta\theta$ along the geodesic trajectory $\Sigma$, followed by projection onto the $S_2$ axis. a)~Properly oriented squeezed ellipsoids allow resolution of the uncertainty regions of the initial and final states. Localization near the pole ($S_{3}=+1$) maximizes the displacement $\delta\langle S_{2}\rangle$. b) Non-oriented squeezed ellipsoids lead to overlap between the uncertainty regions of the initial and final states. In this case, localization near the $S_{3}=+1$ pole alone does not resolve the uncertainty regions.}
    \label{fig:toOptimalMeasurement}
\end{figure}

The framework developed in this work applies to physical systems admitting an effective two-level and two-mode description, where state transformations are described by the SU(2) symmetry group. We consider quantum states described by SU(2) generators $\hat S_{\rm i}$ satisfying the commutation relations $[\hat S_{\rm i},\hat S_{\rm j}] = i \epsilon_{\rm ijk} \hat S_{\rm k}$, and the corresponding uncertainty relation $\Delta \hat{S}_{\rm i}\, \Delta \hat{S}_{\rm j}\geq|\langle \hat{S}_{\rm k}\rangle|/2$. Since the SU(2) group is the double cover of the rotation group SO(3), the mean state of such a system can be mapped onto a point on the surface of a unit sphere. A quantum coherent state can be represented as a small spherical uncertainty distribution centered at the mean-state point. Squeezing deforms this distribution into a three-dimensional uncertainty ellipsoid. In this work, we restrict ourselves to the particular case in which the minor axis of the uncertainty ellipsoid lies in the tangent plane to the sphere. In the local tangential approximation, the uncertainty distribution is then represented by an ellipse whose orientation provides an additional degree of freedom beyond the state position on the sphere and determines the achievable measurement sensitivity, as illustrated in Fig.~\ref{fig:toOptimalMeasurement}.

We now specify the measurement configuration used to define optimality for a small phase shift $\delta\theta$. The phase shift is introduced by a unitary transformation that drives the state along a geodesic trajectory $\Sigma$ on the sphere (see Fig.~\ref{fig:toOptimalMeasurement}), prior to projection onto the eigenbasis associated with the measurement operator defined along the fixed axis. For the sake of transparent geometric interpretation, we focus on a particular configuration of the problem, while the strategy developed here remains general and applicable to arbitrary state positions and measurement geometries.

\paragraph*{Optimal measurement.}
The chosen configuration assumes that phase encoding is generated by rotations about the $S_{1}$ axis,
$\ket{\psi(\delta\theta)} = e^{- i \delta\theta\,\hat S_{1}} \ket{\psi_0}$, and a projection axis $S_{2}$. The phase uncertainty obtained from standard error propagation then reads $\Delta \theta = \Delta \hat S_{2}/| \partial_\theta \langle \hat S_{2} \rangle|,$
which, using the commutation relation given above, yields
$\Delta \theta = \Delta \hat S_{2}/|\langle \hat S_{3} \rangle|$~\cite{Wineland1992,Giovannetti2004,Ma2011,Pezze2018}.
Accordingly, the denominator reaches its maximal value for states localized near either of the two $S_{3}$ poles, corresponding to coordinates $\{0,0,\pm1\}$, where small phase perturbations produce maximal displacement of the expectation value of the measurement operator. In turn, the numerator depends solely on the uncertainty shape.
For squeezed states, minimizing the phase uncertainty reduces to minimizing $\Delta \hat S_{2}$, which is achieved by an appropriate choice of the squeezing ellipse orientation. Thus, in the prescribed configuration, a state ready for optimal measurement combines optimal localization near the $S_{3}$ poles with an orientation of the squeezing ellipse such that its minor axis is aligned along the measurement axis $S_{2}$. The optimization problem therefore extends beyond positioning the mean-state vector and additionally requires proper alignment of the squeezing ellipse.

\begin{figure}[t!]
    \centering
    \includegraphics[width=0.95\linewidth]{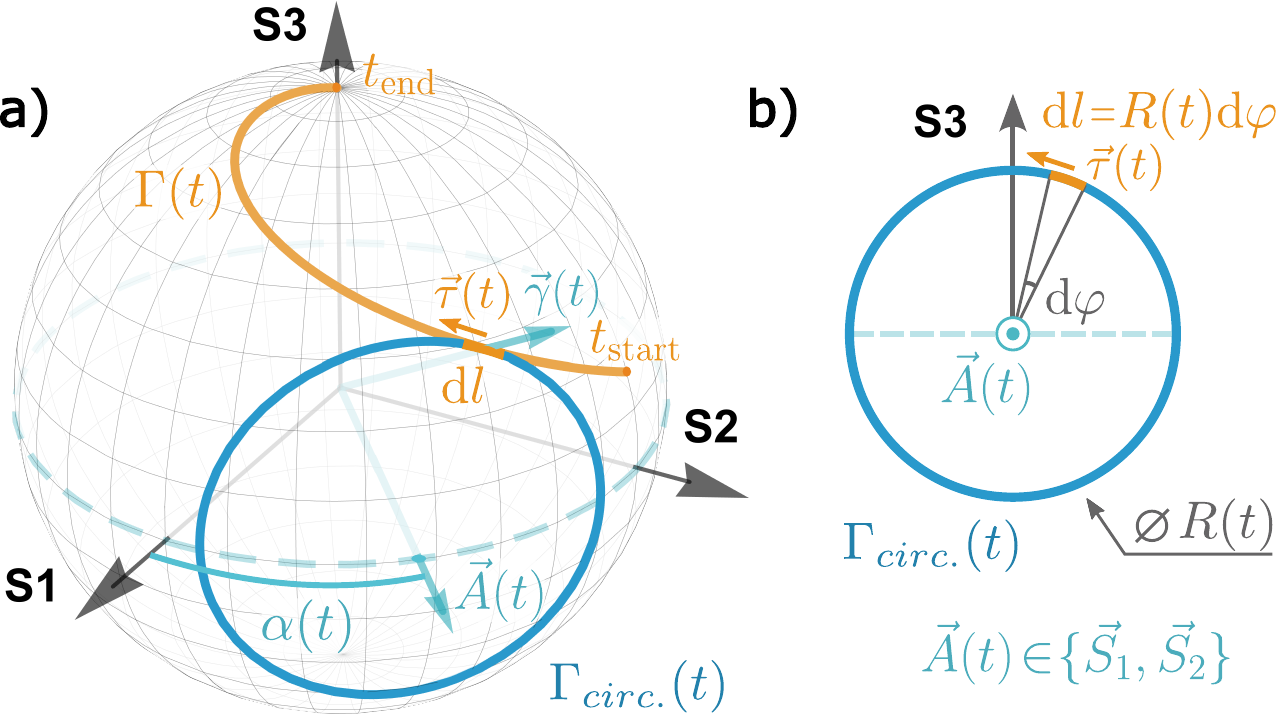}
    \caption{\textbf{Illustrative representation of a smooth curve $\Gamma(t)$ through successive infinitesimal elements of rotation circles.} (a) Coincidence between $\Gamma(t)$ and $\Gamma_{\rm circ.}(t)$ over a local infinitesimal arc element $\dd l$. (b) The arc $\dd l$ represented via a circle constructed about the axis $\vec{A}(t)$, reducing the problem to 2D. $\vec{\tau}(t)$ -- tangent vector to $\Gamma(t)$ shared by both the tangent plane to the sphere and the plane of the auxiliary circle $\Gamma_{\mathrm{circ.}}(t)$. The vector $\vec{A}(t)$ defines the orientation of the waveplate, and lies in the plane $\{S_{1},\,S_{2}\}$. The radius $R(t)$ of the circle $\Gamma_{\rm circ.}(t)$ is determined by geometry of the described system.}
    \label{fig:sketch}
\end{figure}

\paragraph*{Generator representation.}
To formalize the preparation strategy, we consider a unitary operator 
$\hat{U}(t_{\rm end},t_{\rm start})$ that prepares the optimal measurement 
state from an arbitrary initial configuration. The evolution generated by this operator can be geometrically represented by a smooth spherical curve $\Gamma(\theta(t),\phi(t))$ with the corresponding vector $\vec{\gamma}(t)$ specifying the instantaneous mean-state position, see Fig.~\ref{fig:sketch}~(a). The curve lies on the surface of a unit sphere, and its infinitesimal length element $\dd l$, defined at a given set of angular coordinates $(\theta,\,\phi)$, can be expressed as (see, e.g.,~Ch.~3 in~\cite{Lipschutz1969})
\begin{equation}
\dd l_{\Gamma_{\rm sph.}} =|\vec{\tau}(t)|\:\dd t=\sqrt{\dot{\theta}^{2}(t)+\sin^2\theta\,\,{\dot\phi}^{2}(t)}\:\dd t\,.
\label{eq:dLSphere}
\end{equation}
Here $\vec{\tau}=\dot{\vec{\gamma}}$ is the tangent vector, $\dd l_{\Gamma_{\rm sph.}} \in \Gamma$, and we parameterize the curve $\Gamma(\theta,\phi)$ using a single coordinate $t$. An overdot denotes the total derivative with respect to $t$. This parameter defines the angular coordinates $\theta(t)$ and $\phi(t)$, and spans the interval $t \in [0,1]$, representing the beginning and the end of the transformation.

Equivalently, the operator $\hat{U}(t_{\rm end},t_{\rm start})$ can be 
written as a time-ordered exponential of infinitesimal generators 
(see, e.g.,~Sec.~2.1 in~\cite{Sakurai2011}, Sec.~III.F 
in~\cite{CohenTannoudji2020}, and Sec.~4.2 in~\cite{Steck2020}), whereby each 
infinitesimal factor acts locally along the curve $\Gamma(t)$. Accordingly, 
each arc element $\dd l$ may be viewed as part of a spherical circle whose center lies on the axis $\vec A(t)$, given by the intersection of the midpoint plane normal to $\dd l$ with the projective plane $\{S_{1},\,S_{2}\}$ (see Fig.~\ref{fig:sketch}~(b)). This construction allows one to represent the transformation as a rotation by an infinitesimal angle $\dd \varphi$, yielding the identity
\begin{equation}
\dd l_{\Gamma_{\rm circ.}} = R(t)\,\dd \varphi\,,
\label{eq:dLcircle}
\end{equation}
where $R(t)$ is the radius of the given circle and $\dd l_{\Gamma_{\rm circ.}}\in \Gamma_{\rm circ.}(t)$. This shows that any infinitesimal element $\dd l$ corresponds to an SO(3) rotation $\exp\!\left(i\,\mathbb{R}_{\vec{A}}(t)\,\dd\varphi\right)$, where $\mathbb{R}_{\vec{A}}(t)$ is the generator of rotation about the axis $\vec{A}(t)$. This construction is analogous to the action of a time-dependent Hamiltonian over an infinitesimal interval $[t,t+\dd t]$, written as $\exp\!\left(-i\hat{H}(t)\,\dd t\right)$, and implies that a continuous transformation may be represented as a sequence of elementary rotations generated by SU(2) operators. Owing to the double-cover relation between SU(2) and SO(3), each such rotation is generated by a corresponding SU(2) operator acting on the quantum state. With rotation axes limited to a single plane, i.~e., in case of linear birefringent elements, this establishes a direct correspondence between the geometric trajectory on the sphere and the unitary evolution in Hilbert space. Given this representation of infinitesimal rotations, we can proceed to the discussion of the practical concept scheme, which reduces to two primary steps: 1) establishing the trajectory $\Gamma(t)$, and 2) designing an experimentally viable implementation.

\paragraph*{Vector transport rules.} Establishing the transformation trajectory $\Gamma(t)$ requires accounting for the evolution of tangent-plane directions under a continuously varying rotation axis $\vec A(t)$. An infinitesimal rotation changes any tangent-plane vector $\vec{\rm x}(t)$ by an increment $\dd\vec{\rm x}(t)=\vec A(t)\times\vec{\rm x}(t)\,\dd\varphi$. Accordingly, the squeezing ellipse major axis $\vec p_{\rm sq}(t)$ follows this rotation, whereas the evolution of the trajectory tangent vector $\vec{\tau}(t)=\vec{\omega}(t)\times\vec{\gamma}(t)$ also depends on the time variation of the instantaneous angular velocity $\vec{\omega}(t)=\dot{\varphi} \vec{A}(t)$. The corresponding evolution equations read
\begin{align}
\dot{\vec{\tau}}(t)&=\big[\vec{\omega}(t)\times\vec{\tau}(t)\big]+\big[\dot{\vec{\omega}}(t)\times\vec{\gamma}(t)\big]\,,\label{eq:tauEQ} \\ 
\dot{\vec{p}}_{\rm sq}(t)&= \big[\vec{\omega}(t)\times\vec{p}_{\rm sq}(t)\big]\,\label{eq:thetaSQ}.
\end{align}
While the common rotation terms in both equations preserve the relative orientation of the two vectors, the additional term in eq.~\eqref{eq:tauEQ} includes a contribution proportional to $\dot{\vec A}(t)$, which may alter the tangent direction relative to the transported ellipse axis. Consequently, the relative angle between $\vec\tau(t)$ and $\vec p_{\rm sq}(t)$ need not remain constant, despite the angle-preserving nature of each infinitesimal SO(3) rotation.

\begin{figure}[t!]
    \centering
    \includegraphics[width=0.95\linewidth]{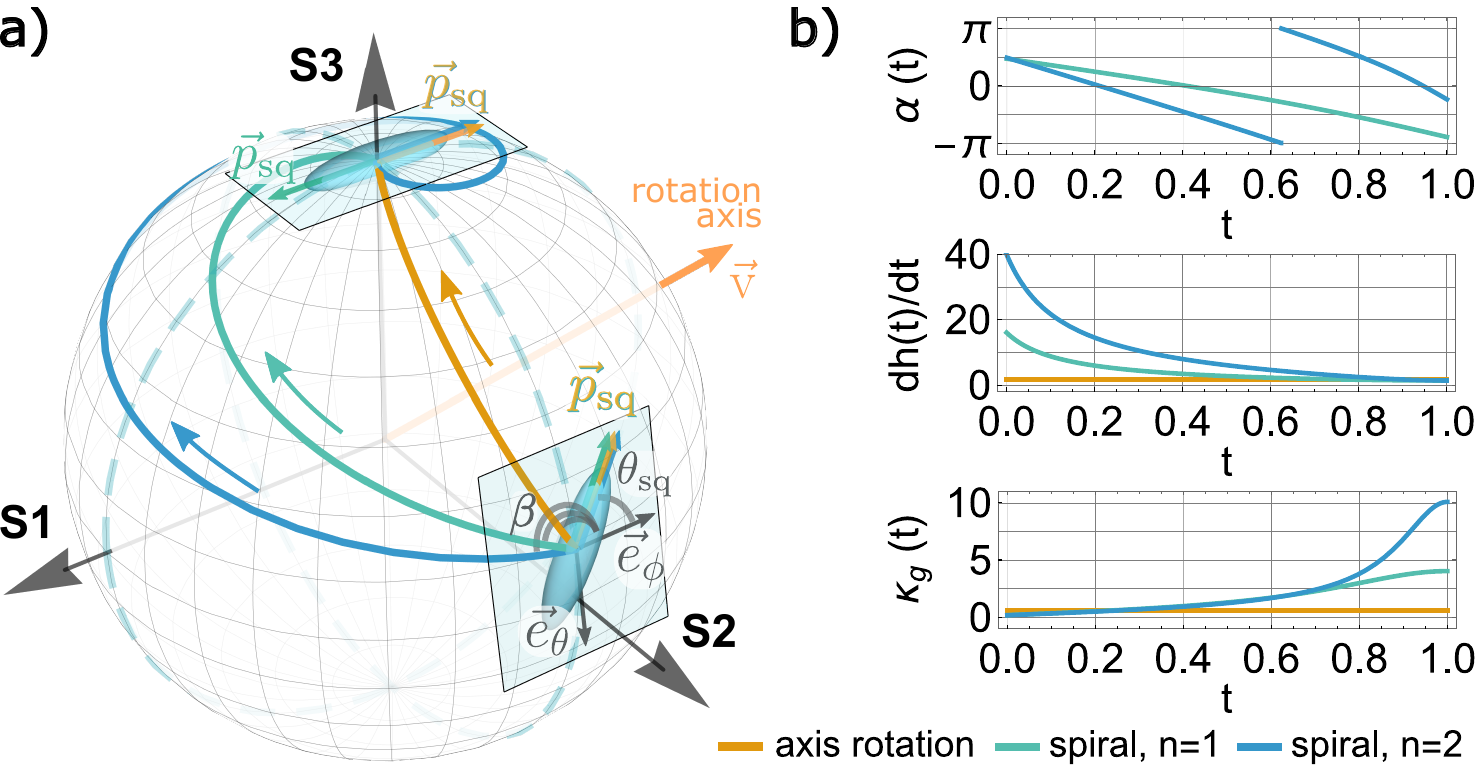}
     \caption{\textbf{Representative transformation trajectories $\Gamma(t)$ and corresponding control parameters.} a) The spiral-like (teal for $n=1$ and blue for $n=2$) and axial-rotation (orange) transformation trajectories. The angle $\theta_{\rm sq}$ specifies the orientation of the squeezing ellipse major axis, represented by the vector $\vec{p}_{\rm sq}$ (colored accordingly) in the tangent plane, relative to the azimuthal direction $\vec{e}_{\phi}$. b) Corresponding control parameters of the multilayer waveplate device. The local layer thickness $\dd h(t)/\dd t$ is obtained from eq.~\eqref{eq:dH} and expressed in units of $\big[\frac{\lambda}{2 \pi (n_{\rm e}-n_{\rm o})}\big]$. The angle $\alpha(t)$ between the axis $S_{1}$ and the vector $\vec{A}(t)$ (see Fig.~\ref{fig:sketch}~(a)) determines the experimentally controlled waveplate orientation given by $\alpha(t)/2$. This relation applies to equatorial-axis transformations and not to the fixed-axis rotation about $\vec{\rm v}$, whose total rotation angle is approximately $97^{\circ}$ in the example shown. The geodesic curvature $\kappa_{\rm g}(t)$ provides auxiliary geometric information characterizing the transformation and equals approximately 0.6 for the $\vec{\rm v}$-axis rotation.}
    \label{fig:spiral}
\end{figure}

According to the optimality condition introduced above, the transformation trajectory $\Gamma(t)$ must have its endpoint $\Gamma(1)$ at the pole while satisfying the requirement of orthogonality between the squeezing major axis $\vec{p}_{\rm sq}(1)$ and the projection axis $S_{2}$, see Fig.~\ref{fig:spiral}~(a). To characterize vector transport along $\Gamma(t)$, we first describe the evolution of the tangent vector $\vec\tau(t)$ by its oriented angle $\beta(t)$ relative to the local azimuthal vector $\vec {e}_{\phi}(t)$, such that $\cos\beta(t)=\widehat{\vec{\tau}}(t)\cdot\vec{e}_{\phi}(t)$, where $\widehat{\vec\tau}(t)$ is the unit tangent vector. For the given initial state, the problem falls within the framework of differential geometry and admits infinitely many trajectories connecting the initial point to the pole, with different initial tangent orientations $\beta(0)$. To determine the direction from which the tangent vector $\vec{\tau}(t)$ approaches the pole relative to the positive $S_{1}$ axis, we apply the Gauss–Bonnet theorem. For simple trajectories whose initial and final tangent vectors lie within the same quadrant of the $\{S_{1},\,S_{2}\}$ plane, so that no winding around the pole occurs, the general theorem formulation reduces to (see, e.g., Ch.~{11} in \cite{Lipschutz1969})
\begin{equation}
\angle\big(\vec{S}_{1},\vec{\tau}(1)\big)
 = \pi+\beta(0) - \int_{\Gamma(t)} \kappa_{\rm g}(t)\, \dd l(t)+\Omega_{\Gamma\cup\Sigma}\,.
    \label{eq:Tsq}
\end{equation}
Here, $\kappa_{\rm g}=(\widehat{\vec{\tau}}\times\vec{\gamma})\cdot\dd\widehat{\vec\tau}/ \dd l$ is the geodesic curvature, and $\Omega_{\Gamma\cup\Sigma}$ is the solid angle swept by the closed piecewise smooth contour formed by $\Gamma(t)$ and an auxiliary geodesic arc $\Sigma$ connecting its endpoints. Here, the auxiliary geodesic arc has zero curvature $\kappa_{\rm g}=0$ and thus gives no contribution to the contour integration (see, e.g., \cite{Samuel1988,Bhandari1991,Bliokh2019,Zhou2020,Leinonen2023}).

The corresponding local dynamic evolution of $\beta(t)$ relative to the azimuthal vector $\vec{e}_{\phi}(t)$ along $\Gamma(t)$ can be expressed as follows: $\dot{\beta}(t)=-\kappa_{\rm g}(t)\,\dot{l}(t)-\cos\theta(t)\,\dot\phi(t)\,.$ The second term describes the rotation of the local azimuthal basis along the trajectory and connects the tangent evolution to the geometric phase, which is proportional to the solid-angle integral $\int_{\Gamma\cup\Sigma}(1-\cos\theta)\,d\phi$ (see, e.g.,~\cite{Cisowski2022}). In this way, shaping the trajectory $\Gamma(t)$ by prescribing the evolution of its tangent direction allows one to control the accumulated geometric-phase contribution, subject to the prescribed boundary conditions. 

Having discussed the tangent evolution and its connection to the geometric phase, we turn to the orientation of the squeezing ellipse. In the measurement configuration considered here, the required final orientation at the pole satisfies $\angle (\vec{p}_{\rm sq}(1),\vec{S_{1}})=\pi n$, where $n\in \mathbb{Z}$. To impose this condition, we introduce the oriented squeezing angle $\theta_{\rm sq}(t)$ between the major axis $\vec{p}_{\rm sq}(t)$ and the local azimuthal vector $\vec{e}_{\phi}(t)$, using the same orientation convention as for $\beta(t)$. In the general case, where the rotation axis $\vec{A}(t)$ is not restricted to the $\{S_{1},\,S_{2}\}$ plane, the vector transport rule introduced in eq.~\eqref{eq:thetaSQ} gives
\begin{equation}
\dot\theta_{\rm sq}(t)=\vec{\omega}(t)\cdot\vec{\gamma}(t)-\cos\theta(t)\dot{\phi}(t)\,.
\end{equation}
Here, the scalar product in the first term accounts for the radial component of the angular velocity, while the local-basis term retains its connection with the solid-angle integral discussed above. For the configuration considered in this work, with the rotation axis confined to the equatorial plane, $\vec{A}(t)\in\{S_{1},S_{2}\}$, the expression drastically simplifies to
\begin{equation}
\dot\theta_{\rm sq}(t)=-\frac{\dot{\phi}(t)}{\cos\theta(t)}.
\label{eq:dotThetaSq}
\end{equation}
This equation applies to trajectories confined to either open hemisphere, $\theta(t)\ne\pi/2$, and considerably simplifies the construction of $\Gamma(t)$ under the prescribed boundary conditions.

Thus, for a given initial squeezing angle $\theta_{\rm sq}(0)$, the choice of trajectory $\Gamma(t)$ determines the accumulated ellipse rotation and allows the required alignment at the pole to be achieved. Although the accumulated ellipse rotation differs from the geometric phase, its local-basis contribution is associated with the same solid-angle integral that enters the geometric-phase description. The proportionality factor relating the geometric phase to the solid angle depends on the specific physical system (see, e.g.,~\cite{Cisowski2022}), while the evolution of the local spherical basis provides a common geometric link between the geometric-phase accumulation and the rotation of the squeezing ellipse.

This correspondence leads to the \emph{second step} of the construction, namely selecting a physical implementation reproducing the action of the generator ${\rm exp}\!\left(i\,\mathbb{R}_{\vec{A}}(t)\,\dd\varphi\right)$ on the state vector within a given physical setting. In this context, suitable realizations can be drawn from techniques used for geometric-phase control. Possible physical implementations of the proposed scheme for preparing an optimal measurement state will be discussed in the following section.

\section*{Implementation Schemes} 
\paragraph*{Poincar\'e sphere.}
To illustrate the construction, we consider polarization states mapped onto the Poincar\'e  sphere. This platform provides a simple geometric picture while retaining the generality of the approach. In this representation, each rotation by an angle $\dd\varphi$ associated with the element $\dd l_{\rm sph.}$ is equivalent to the action of a birefringent waveplate retarder on the polarization state, see, e.g., Ch.~5 in ~\cite{Chekhova2021} and Ch.~7 in \cite{Shurcliff1962}. Accordingly, the transformation $\Gamma(t)$ is realized as a sequence of waveplates whose optical-axis orientation and thickness vary smoothly with $t$. Half of the spherical azimuthal angle $\alpha(t)$, the angle which determines the direction of $\vec{A}$ on the plane $\{S_{1},\, S_{2}\}$, corresponds to the physical orientation of the ordinary axis of the waveplate retarder. The angle $\alpha(t)$ is determined geometrically. Specifically, it is the azimuthal angle of the vector orthogonal to the projection of the tangent vector $\vec{\tau}(t)$ onto the plane $\{S_{1},\, S_{2}\}$, yielding $\alpha(t) = \mathrm{atan2}\!\left(\tau_{S_{1}}(t),\,-\tau_{S_{2}}(t)\right)$. The waveplate thickness $h$ sets the rotation angle about $\vec{A}$ according to expression $\varphi=2 \pi (n_{\rm e}-n_{\rm o}) h /\lambda$, where $n_{\rm o}$ and $n_{\rm e}$ are the refractive indices of the ordinary and extraordinary axes of the plate, respectively, and $\lambda$ is the wavelength of the squeezed radiation field. Substituting the total differential of this expression into eq.~\eqref{eq:dLcircle}, and using the equivalence of the arc elements $\dd l_{\Gamma_{\rm sph.}}$ in eq.~\eqref{eq:dLSphere} and $\dd l_{\Gamma_{\rm circ.}}$, we obtain the local thickness $\dd h (t)$ of the layer associated with the element $\dd l_{\Gamma_{\rm sph.}}$ at the parameter $t$ as follows:
\begin{equation}
    \dd h (t)=\frac{\lambda}{2 \pi (n_{\rm e}-n_{\rm o}) R(t)} \dd l_{\Gamma_{\rm sph.}}\,.
    \label{eq:dH}
\end{equation}
As a result, this equation and the relation for $\alpha(t)$ uniquely determine the spatial structure of the birefringent element implementing the transformation $\Gamma(t)$. The correspondence between the dimensionless trajectory parameter $t$ and the physical time $T$ is established through the transit time across each infinitesimal layer: $dT=(n_{\rm o}/c)\,\dd h$, thereby setting the physical time scale of the transformation.

The finite propagation speed of light in the birefringent medium connects the spatial structure of the element to the temporal evolution of the state. Practical constraints on this evolution may therefore limit the physically realizable parameter profiles. Because $\Gamma(t)$ is only constrained by smoothness and the prescribed boundary conditions, its shape can be tailored to the capabilities of an experimental setup to control these parameters with the required precision.

To examine this design freedom, we consider two representative classes of transformation trajectories, highlighting distinct implementation regimes and control over the birefringent-element parameters. The initial position of the squeezing ellipse in spherical coordinates is chosen to be at $ \left\{1,\,\theta_{0},\,\pi/2\right\}$ with squeezing angle $\theta_{\rm sq}$, see Fig.~\ref{fig:spiral}~(a), corresponding to experimentally relevant conditions reported in~\cite{Kalinin2023,Kalinin2023a}.

The \textit{first transformation} is a circular arc of angular length $\Phi$. It is given by a single rotation ${\rm exp}\!\left(i\,\mathbb{R}_{\vec{\mathrm{v}}}\,\Phi\right)$ about a fixed vector $\vec{\mathrm{v}}$ determined by the initial conditions. This example demonstrates the feasibility of implementing the alignment transformation using a single experimentally available optical element. With the rotation axis fixed, $\dd\vec{\mathrm{v}}/\dd t=0$, the resulting rigid rotation preserves the angle between the major axis of the squeezing ellipse $\vec{p}_{\rm sq}$ and $\vec{\tau}$. For generic initial conditions, $\vec{\rm v}$ lies outside the equatorial $\{S_{1},\,S_{2}\}$ plane (${\rm v}_{S3}\neq0$), corresponding in the Poincar\'e-sphere representation to a wave plate with elliptical birefringence, which can be realized using freeform metasurfaces (see, e.g.,~\cite{Shi2020}).

The \textit{second transformation} corresponds to a spiral trajectory winding about the pole, with an integer index $n$ labeling the branches of the final alignment condition and is parametrized as follows: 
\begin{align}
\theta(t)= \theta_{0}-\upsilon_{\theta}\,t\,\,\,\,\,\,\,\,\,{\rm and}\,\,\,\,\,\,\,\,\,\phi(t) = \frac{\pi}{2}-\upsilon_{\phi,n}\,t\,\,\,,
\label{eq:spiralEQS}
\end{align}
where $\upsilon_{\theta}$ and $\upsilon_{\phi,n}$ are the constant polar and azimuthal speeds, respectively, determined by the boundary conditions at $t=0$ and $t=1$. To account for the azimuthal ambiguity at the pole, we determine the limiting orientation of the local spherical basis along the prescribed trajectory by defining $\vec{e}_{\phi}(1)=\{-\sin\phi(1),\,\cos\phi(1),\,0\}$. Integrating eq.~\eqref{eq:dotThetaSq} along the spiral trajectory defined by eqs.~\eqref{eq:spiralEQS} and imposing the endpoint alignment condition yields an explicit expression for $\upsilon_{\phi,n}$:
\begin{equation}
\upsilon_{\phi,n}=\frac{\pi n -\theta_{\rm sq}(0)}{\arctanh(\sin \upsilon_{\theta})/\upsilon_{\theta}-1}\,, \quad {\rm where}\quad n\in \mathbb{Z}\,.
\end{equation}
For the illustrative example shown in Fig.~\ref{fig:spiral}(a), with initial conditions $\theta_0=80^\circ$ and $\theta_{\rm sq}(0)=60^\circ$, we set $n=1$ and $n=2$, chosen to demonstrate the freedom in selecting the geodesic curvature along the transformation.

The freedom in trajectory design allows the profiles $\alpha(t)$ and $\mathrm{d}h(t)/\mathrm{d}t$ to be chosen subject to additional experimental requirements. A natural physical implementation of the proposed transformation is provided by liquid-crystal elements with a dynamically varying director orientation, see, e.g., \cite{Sit2024a,Sit2024b}. In the nematic phase, the optical axis can be continuously rotated along the propagation direction by an external electric field, enabling a smooth implementation of the trajectory $\Gamma(t)$.

In this sense, the system is analogous to the curved optical beam considered by Rytov, for which Vladimirskii showed that the transport of the local frame along the ray leads to an accumulated polarization rotation equal to the solid angle $\Omega$ enclosed by the trajectory, see~\cite{Rytov1938,Vladimirskii1941,Vinitskii1990,Berry1990}. The experiment was later proposed and confirmed in optical systems~\cite{Chiao1986,Tomita1986}.

\begin{figure}[t!]
    \centering
    \includegraphics[width=0.95\linewidth]{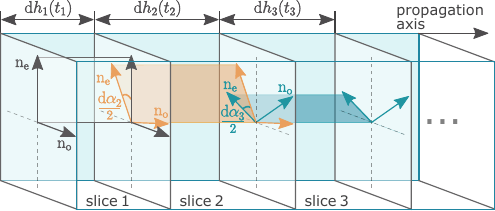}
    \caption{\textbf{Multilayer waveplate stack approximating the continuous transformation.} The thickness of the layer $\dd h_{\rm i}(t_{\rm i})$ is determined by eq.~\eqref{eq:dH}. The orientation of the waveplate axis to the vertical-horizontal basis is half the angle between axis $S_{1}$ and vector $\vec{A}(t)$ shown in Fig.~\ref{fig:sketch}.}
    \label{fig:stack}
\end{figure}

At a more practical level, the continuous transformation can be approximated by a stack of birefringent waveplate retarders, see Fig.~\ref{fig:stack}. Concatenations of discrete waveplates have been widely considered for geometric-phase and polarization control (see, e.g.,~\cite{Berry1996,Shore2016,Alessandrini2025}). In the limiting case, a minimal set of three waveplates, known as the Simon-Mukunda polarization gadget~\cite{Simon1990}, is sufficient to perform any possible rotation on the Poincar\'e sphere, i.~e., bring a polarization-squeezed state to any desired mean-state position with arbitrary squeezing-ellipse alignment in the local tangent plane. In experiments with polarization-squeezed light, such control was used for setting the squeezing ellipse of a prepared squeezed state to the optimal measurement orientation (see, e.g., \cite{Kalinin2023,Kalinin2023a}). However, such a discrete implementation replaces the continuous evolution along the transformation curve $\Gamma(t)$ with a sequence of stepwise operations, thereby eliminating the possibility of dynamical control of the transformation along the parameter $t$.

This discretization raises the question of how accurately a finite sequence of waveplates can reproduce a smooth geometric trajectory $\Gamma(t)$. Approximating the continuous evolution by a succession of circular arcs $\Delta \Gamma_{{\rm circ.},{\rm i}}$, such that $\Gamma(t)\approx\bigcup_i\Delta \Gamma_{{\rm circ.},{\rm i}}$, generally changes the intermediate trajectory, even when the desired final mean-state position is reproduced. Consequently, equidistant arc-length steps do not generally ensure uniform approximation accuracy, since variations in geodesic curvature $\kappa_g$ can lead to rapid changes in the waveplate-axis orientation $\alpha(t)$, requiring non-uniform discretization to maintain sufficient local resolution. These limitations motivate continuous implementations in applications where the intermediate quantum-state evolution is relevant, while the squeezing ellipse must attain the prescribed orientation at the endpoint.

\paragraph*{Bloch sphere.} The proposed procedure is not restricted to polarization optics and applies to systems described by SU(2) dynamics and SO(3) rotations, whose quantum fluctuations may exhibit squeezing. Examples include collective pseudospin systems in atomic or solid-state ensembles of two-level atoms, which support squeezed collective spin states (see, e.g.,~\cite{Wineland1992, Kitagawa1993, Ma2011, Sinatra2022}). Related scenarios also arise in platforms relevant to wave-matter interferometry. A characteristic realization of the desired transformation $\Gamma(t)$ can be achieved using a variety of established control techniques. These include precise control of an external driving field via continuous modulation of the Rabi frequency, as well as composite pulse sequences (see, e.g., \cite{Levitt1986,Torosov2019}). In particular, a physically transparent realization is provided by systems of real spins, where the transformation corresponds to explicit control of the spin orientation in physical space. In this case, the desired trajectory $\Gamma(t)$ can be implemented by tailoring the magnitude and direction of an externally applied magnetic field over a given time interval, thereby inducing controlled spin precession at the Larmor frequency. Taken together, these considerations indicate that the geometric control strategy developed here can be implemented in a wide class of atomic and solid-state systems, providing a route toward experimental realization of the proposed transformation.

\paragraph*{Orbital Poincar\'e sphere.}
Another relevant class of systems is formed by spatial optical modes mapped onto the orbital Poincar\'e  sphere (see, e.g.,~\cite{Padgett1999,Dennis2017,Shen2020}), such as Hermite-Gaussian ($\rm HG_{10},HG_{01}$) and Laguerre-Gaussian ($\rm LG_{01}$) modes. Squeezing in such spatial-mode systems has been demonstrated experimentally (see, e.g.,~\cite{Lassen2009}). In this setting, applying the unitary transformation described above would require independent control at each transverse spatial coordinate, implying that the transformation must act locally across the entire beam profile. A fully continuous physical implementation of this kind remains restricted by current experimental capabilities. Nevertheless, recent interest in converters for higher-order Poincar\'e  spheres (see, e.g.,~\cite{Marco2022,Bansal2025,Umar2026}) suggests that more complete control over transformations on the orbital Poincar\'e  sphere may become accessible in the near future.

Despite these limitations, certain transformations on the orbital Poincar\'e  sphere can already be realized. Partial implementations are provided by cylindrical lens systems (see, e.g.,~\cite{Beijersbergen1993,Galvez2003}) and spatial light modulators (see, e.g.,~\cite{Volyar2024}). Discrete implementations based on stacks of spatial phase elements, such as multi-plane light converters (see, e.g.,~\cite{Fontaine2019,MartinezBecerril2024}), offer an alternative route via piecewise approximation of the desired transformation. However, the practical feasibility and fidelity of such discretized realizations, including approximations of $\Gamma(t)$ by circular arc segments $\Delta \Gamma_{{\rm circ.},{\rm i}}$, remain open problems for future investigation.

\section*{Conclusion}
We have introduced a geometric framework for preparing optimal measurement states from arbitrary squeezed inputs in SU(2)-symmetric systems. In particular, we have illustrated how a succession of SO(3) rotations moves the mean-state position along a continuous trajectory on the unit sphere while varying the orientation of the squeezing ellipse. The resulting change in the ellipse's orientation includes a contribution from the accumulated rotation of the local tangent-plane basis, which is also accounted for in the geometric-phase description.

This extends the geometric description to include the evolution and control of the squeezing orientation, together with the accompanying geometric-phase variation, making the choice of transformation trajectory a practical tool for metrological state preparation. A suitable trajectory can thus be designed to simultaneously realize the desired state transformation and establish the squeezing orientation required for optimal measurement, thereby enhancing the achievable sensitivity. Such a trajectory-based approach holds promise as an auxiliary tool for preparing squeezed states with orientations tailored to a prescribed subsequent evolution, a capability that could support experimental studies of quantum speed limits (see, e.g.,~\cite{Pires2016, Xu2019a}) and geometric-phase evolution in open systems (see, e.g.,~\cite{Carollo2003}). For polarization-squeezed light, the framework yields an explicit implementation in terms of continuously varying birefringent elements, with waveplate stacks as discrete approximations. More broadly, the same geometric principle applies to other platforms combining SU(2) control with squeezing, providing a unified route toward optimized quantum measurements.

We believe that the illustrative geometric picture discussed here will foster physical intuition and facilitate the design and interpretation of experiments involving squeezing-orientation control for metrological state preparation.

\section*{Acknowledgments}
The authors acknowledge colleagues at the Max Planck Institute for the Science of Light for drawing our attention to this research direction, from which the present work originated, and for helpful comments and reference suggestions. V.S. acknowledges useful discussions and helpful communications with Sergey Kochubei and thanks Sharareh Sayyad for valuable comments during the preparation of the manuscript.

\bibliography{bibfile}
\end{document}